# Weakly anisotropic superconductivity of Pr$_4$Ni$_3$O$_{10}$ single crystals

Cuiying Pei [a§], Yang Shen [a§], Di Peng [b,c§], Mingxin Zhang [a,d§], Yi Zhao [a], Xiangzhuo Xing [e], Qi Wang [a,f], Juefei Wu [a], Junjie Wang [a], Lingxiao Zhao [a], Zhenfang Xing [b,c], Yulin Chen [a,f,g], Jinkui Zhao [d], Wenge Yang [b,c], Xiaobing Liu [e], Zhixiang Shi [h], Hanjie Guo [d*], Qiaoshi Zeng [b,c*], Guang-Ming Zhang [a,i*] and Yanpeng Qi [a,f,j*]

[a] *State Key Laboratory of Quantum Functional Materials, School of Physical Science and Technology, ShanghaiTech University, Shanghai 201210, China*
[b] *Center for High Pressure Science and Technology Advanced Research, Shanghai 201203, China*
[c] *Shanghai Key Laboratory of Material Frontiers Research in Extreme Environments (MFree), Institute for Shanghai Advanced Research in Physical Sciences (SHARPS), Shanghai 201203, China*
[d] *Songshan Lake Materials Laboratory, Dongguan 523808, Guangdong, China*
[e] *Laboratory of High Pressure Physics and Material Science, School of Physics and Physical Engineering, Qufu Normal University, Qufu 273165, Shandong, China*
[f] *ShanghaiTech Laboratory for Topological Physics, ShanghaiTech University, Shanghai 201210, China*
[g] *Department of Physics, Clarendon Laboratory, University of Oxford, Parks Road, Oxford OX1 3PU, UK*
[h] *School of Physics, Southeast University, Nanjing 211189, Jiangsu China*
[i] *Department of Physics, Tsinghua University, Beijing 100084, China*
[j] *Shanghai Key Laboratory of High-resolution Electron Microscopy, ShanghaiTech University, Shanghai 201210, China*
* Email: qiyp@shanghaitech.edu.cn
* Email: zhanggm@shanghaitech.edu.cn
* Email: zengqs@hpstar.ac.cn
* Email: hjguo@sslab.org.cn
§ *Cuiying Pei, Yang Shen, Di Peng and Mingxin Zhang contributed equally to this work.*

**ABSTRACT**

Since the discovery of high-temperature superconductivity, studying the upper critical field and its anisotropy has been crucial for understanding superconducting mechanism and guiding applications. Here we perform *in situ* high-pressure angular-dependent electrical transport measurements on $Pr_4Ni_3O_{10}$ single crystals using a custom diamond anvil cell (DAC) rotator and confirming its anisotropic superconductivity. The anisotropy parameter $\gamma$, derived from the upper critical fields ($\mu_0H_{c2}$) for $H \perp ab$ and $H \mathbin{/\mkern-6mu/} ab$, is approximately 1.6, decreasing with increasing temperature and approaches 1 near $T_c$. Comparing effective mass anisotropy and inter-block distance in cuprates and iron-based superconductors (FeSCs) reveals that $Pr_4Ni_3O_{10}$ single crystals superconductors are consistent with a two-band model, where intralayer quantum confinement within the unit cell induces interlayer coherence, thereby leading to three-dimensional (3D) superconductivity. This study not only establishes the existence of anisotropic superconductivity in bulk Ruddlesden-Popper nickelates, but also provide critical insight into the role of dimensionality in high-temperature superconductivity.

**INTRODUCTION**

High-temperature superconductivity has recently been observed in nickelate-based superconductors [1-7], prompting natural comparisons with cuprates and FeSCs. Cuprates, which are characterized by the layered crystal structures and quasi-two-dimensional electronic properties, exhibit pronounced anisotropy in superconducting properties [8-11]. In bilayer and trilayer cuprates, *d*-wave pairing is mediated by interlayer Josephson coupling which enhances superconducting phase coherence, and leads to a substantial enhancement of $T_c$ [12-13]. In contrast, FeSCs exhibit much more 3D characteristics and consequently display rather isotropic superconducting behaviors [14]. The 3D electronic structure, stemming from multi-orbital hybridization and 3D band dispersion, enhances superconducting pairing through multi-band interactions, spin/orbital fluctuation mediation, and suppression of detrimental quantum critical effects.[15] Anisotropy serves as a key parameter indicative of the electronic structure's dimensionality and topology, thereby providing essential insights into the superconducting mechanism and forming a bridge between microscopic orbital physics and macroscopic superconducting behavior. Furthermore, as an important characteristic, the anisotropy parameter and its temperature dependence can directly inform strategies for potential applications.

Nevertheless, studying the anisotropy of bulk crystalline Ruddlesden-Popper (RP) nickelate superconductors faces significant challenges, primarily because their superconductivity only emerges under high pressure. First, the combination of high $T_c$ and strong correlations leads to an extremely large upper critical field. For example, $La_3Ni_2O_7$ exhibits a $T_c$ of 78 K and an estimated zero-temperature upper critical field ($\mu_0H_{c2}(0)$) of ~186 T at 18.9 GPa [1], $La_2PrNi_2O_7$ shows $T_c$ = 82.5 K and $\mu_0H_{c2}(0)$ ~104 T at 16 GPa [3], $La_4Ni_3O_{10}$ reaches $T_c$ = 42.5 K at 42.5 GPa and $\mu_0H_{c2}(0)$ ~48 T at 69 GPa [4-5], and $Pr_4Ni_3O_{10}$ achieves $T_c$ = 40 K and $\mu_0H_{c2}(0)$ ~53 T at 80 GPa [7, 16]. Second, standard commercial Dewar for high-pressure electrical transport measurements lack sufficient bore space to accommodate out-of-plane rotation of the DAC. Although high-magnetic-field facilities can characterize the $\mu_0H_{c2}$ of nickelate superconductors near $T_c$, the values at lower temperatures remain prohibitively high to achieve experimentally. To overcome this limitation, we have developed a novel DAC rotator system housed within a 14 T large-bore Dewar, which enables comprehensive angle-dependent transport measurements under tunable temperature and magnetic field conditions.

Here, we report high-pressure, *in situ* angular-dependent electrical transport measurements on $Pr_4Ni_3O_{10}$ single crystal, performed under a magnetic field rotated along the out-of-plane direction. The superconducting properties of $Pr_4Ni_3O_{10}$ are characterized by a weak anisotropy ($\gamma \sim 1.6$), which exhibits a monotonic decrease with increasing temperature. This anisotropy is notably smaller than that of the typical cuprate superconductor Bi-2212[17] ($\gamma \sim 60$) and the iron-based superconductor $KFe_2Se_2$[18] ($\gamma \sim 6.8$). Furthermore, the correlation between structural and superconducting anisotropy is elucidated from the perspective of electronic structure topology. Notably, despite sharing key structural features with cuprate superconductors—such as a multilayer structure and a quasi-two-dimensional square lattice—$Pr_4Ni_3O_{10}$ exhibits exceptionally small anisotropy. Our findings not only elucidate the anisotropic superconducting behavior in this compound but also provide crucial insights for understanding the superconducting mechanism in nickelates.

**RESULTS AND DISCUSSION**

High-quanlity $Pr_4Ni_3O_{10}$ single crystals were synthesized *via* the vertical optical-image floating zone method (HKZ-300 furnace) at 140-150 bar $pO_2$.[7] In detail, dried $Pr_6O_{11}$ and NiO powders were ground, sintered (1050 °C for 48 h, then 1100 °C for 24 h after pressing into rods), and grown at 5 mm/h with counter-rotated feed/seed rods (27/20 rpm) following a 30 mm/h fast-traveling step. Ambient-pressure characterizations (X-ray diffraction, energy dispersive spectroscopy, and thermogravimetric analysis) confirm the high quality of $Pr_4Ni_3O_{10}$ single crystals with monoclinic crystal structure, target Pr:Ni atomic ratio (~4:3.09), and near-stoichiometric oxygen content (~9.99).[7] $Pr_4Ni_3O_{10}$ single crystals undergo pressure-induced superconducting transition, with zero resistance achieved by optimizing the hydrostatic environment *via* gas pressure-transmitting medium. (Figure S1) Its bulk superconductivity is confirmed with a superconducting volume fraction exceeding 80% by *in situ* high-pressure direct current (d.c.) magnetic susceptibility measurements.[7] *In situ* high-pressure low-temperature synchrotron XRD measurements confirm the pressure-induced structural phase transition in $Pr_4Ni_3O_{10}$, and identify the high-pressure superconducting phase as the tetragonal *I*4/*mmm* structure. (Figure S2)

For electrical transport measurements, four electrodes were fabricated in van der Pauw geometry on single crystals of $Pr_4Ni_3O_{10}$ immersed in a helium gas as pressure transmitting medium (PTM). The electrical current was applied within the *ab*-plane. Upon compression to a pressure of 50.2 GPa, the sample exhibited a superconducting transition with an onset critical temperature ($T_c^{onset}$) of 31 K. To investigate the anisotropy of the superconducting state in $Pr_4Ni_3O_{10}$, a diamond anvil cell (DAC) was mounted on a custom-built rotation stage. The DAC was aligned such that its compression axis was parallel to the applied magnetic field. The sample's rotation center was carefully positioned within the homogeneous region of the magnetic field. This initial orientation was defined as the 0° reference angle. Angular-dependent measurements were subsequently performed by rotating the sample clockwise from this reference in incremental steps.

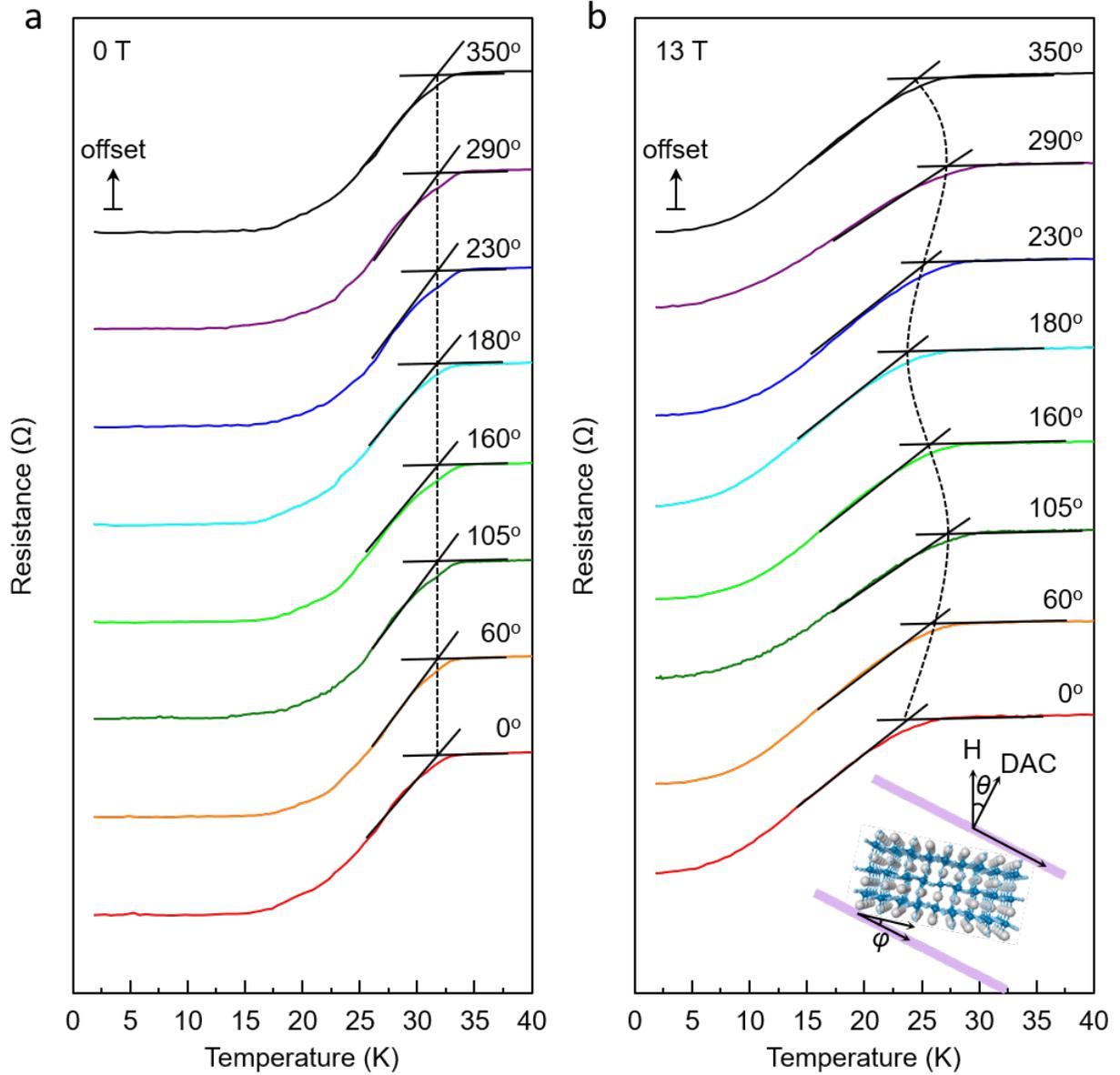

**Figure 1.** Anisotropy in a Pr$_4$Ni$_3$O$_{10}$ single crystal. Temperature dependence of the in-plane resistance ($R_{xx}$) at (a) 0 T and (b) 13 T for various out-of-plane rotation angles ($\theta$). The inset depicts the rotation geometry, where the angle $\varphi$ denotes the tilt between the diamond anvil cell (DAC) culet and the crystalline *ab*-plane. The resistance-temperature curves are vertically offset for clarity. The dashed line serves as a visual guide. The blue, light blue and grey balls represent Ni, O and La atoms, respectively.

The angular dependence of the longitudinal resistivity was investigated by performing out-of-plane rotations of the diamond anvil cell (DAC). Figure 1a presents the temperature-dependent resistance $R(T)$ curves measured at various out-of-plane angles ($\theta$) in the absence of an external magnetic field. Here, the angle $\theta$ is defined as the angle between the compression axis of the DAC and the direction of the applied magnetic field. The $T_c^{onset}$ remains consistent with the value measured prior to the installation of the rotation device and exhibits no angular dependence (Figure S1). Upon application of a magnetic field up to 13 T, the $T_c^{onset}$ progressively shifts to lower temperatures. (Figures S3-S8) This suppression of $T_c^{onset}$ is most pronounced, reaching a minimum, at an angle of 105° as the magnetic field is rotated from out-of-plane to in-plane relative to the *ab*-plane. Further angular tuning of $\theta$ revealed a periodic modulation of the superconducting transition. (Figure 1b) The observed angular dependence of the $T_c^{onset}$ provides direct evidence for superconducting anisotropy in $Pr_4Ni_3O_{10}$ single crystals.

Subsequently, we conducted temperature-dependent resistance measurements at various magnetic fields for different $\theta$. In all cases, the application of a magnetic field resulted in a systematic shift of the superconducting transition to lower temperatures. It is noteworthy that a concurrent broadening of the transition was also observed under applied magnetic fields. Similar behavior has been reported in iron-pnictides [19-21] and chalcogenides [22], as well as cuprate [23], which is commonly attributed to the mechanism of thermally activated flux flow (TAFF). We estimated the upper critical field $\mu_0H_{c2}(T)$ by fitting the experimental data to the Ginzburg-Landau (G-L) formula $\mu_0H_{c2}(T) = \mu_0H_{c2}(0)(1 - t^2)/(1 + t^2)$, where $t = T/T_c$.[24-25] The zero-temperature upper critical field, $\mu_0H_{c2}(0)$, for the $Pr_4Ni_3O_{10}$ single crystals was determined from the field dependence of the temperature at which the resistance reaches 90% of the normal-state value. Notably, the maximum value of $\mu_0H_{c2}(0)$ was not observed at 90°, but rather at 105°, indicating a deviation from the expected magnetic field direction. The 15° deviation likely stems from non-parallelism between the single crystal's *ab*-plane and the diamond culet, as the crystal is suspended in the PTM within the DAC chamber and precise alignment is challenging to achieve during electrodes adjustment.

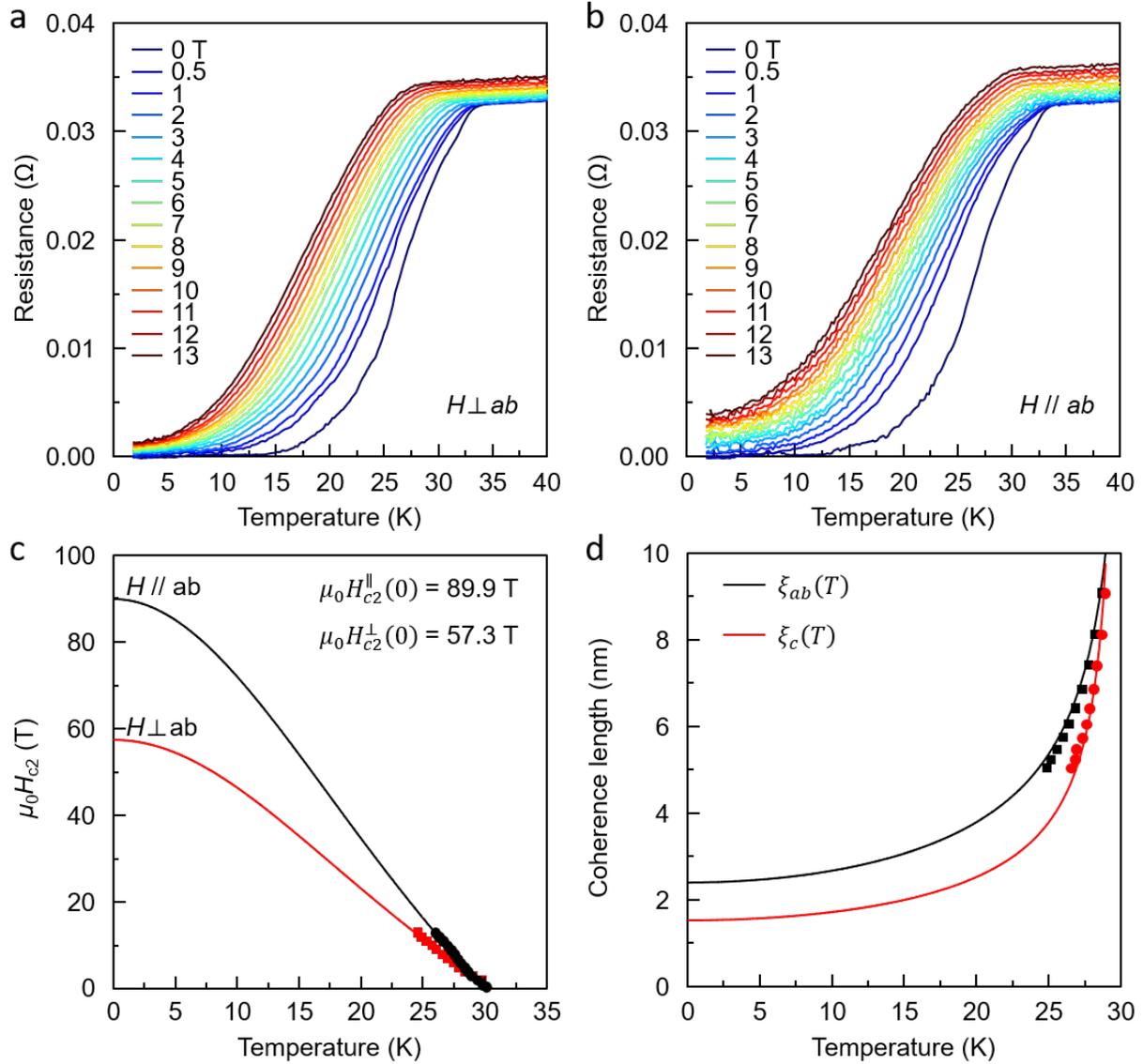

**Figure 2.** Anisotropy of the upper critical field and coherence length in a $Pr_4Ni_3O_{10}$ single crystal. Temperature dependence of the in-plane resistance ($R_{xx}$) under magnetic fields applied perpendicular (H $\perp$ *ab*) (a) and parallel (H//*ab*) (b) to the crystallographic *ab*-plane. (c) Temperature dependence of the upper critical field, $\mu_0 H_{c2}(T)$. $T_c$ is defined at 90% of the normal-state resistance. The solid line is a fit to the G-L model. (d) Temperature evolution of the coherence lengths in the *ab*-plane ($\xi_{ab}(T)$, black squares) and along the *c*-axis (($\xi_c(T)$, red circles), calculated

from the perpendicular ($\mu_0 H_{c2}^\perp(T)$) and parallel ($\mu_0 H_{c2}^\parallel(T)$) critical fields, respectively. The solid lines are derived from the G-L fits in Figure 2c.

We used the $\varphi = 15°$ tilt of the DAC as the out-of-plane reference direction to measure the temperature-dependent resistance of Pr$_4$Ni$_3$O$_{10}$, with magnetic fields applied perpendicular and parallel to the *ab*-plane (Figure 2a and 2b, respectively). From these measurements, the zero-temperature upper critical fields were estimated as $\mu_0 H_{c2}^\parallel(0) = 89.9$ T and $\mu_0 H_{c2}^\perp(0) = 57.3$ T (Figure 2c). The anisotropic ratio, $\gamma$, defined as $\gamma = \mu_0 H_{c2}^\parallel(0)/\mu_0 H_{c2}^\perp(0)$, was determined to be 1.6. The zero-temperature G-L coherence length within the *ab*-plane, $\xi_{ab}(0)$ was calculated to be 2.4 nm using the relation $\mu_0 H_{c2}^\perp = \Phi_0/2\pi\xi_{ab}^2$, where $\Phi_0 = 2.07 \times 10^{-15}$ Wb is the flux quantum. Furthermore, the coherence length along the *c*-axis, $\xi_c(0)$, was estimated to be 1.5 nm *via* the expression $\xi_c(0) = \Phi_0/2\pi\xi_{ab}(0)\mu_0 H_{c2}^\parallel(0)$. As illustrated in Figure 2d, the coherence lengths along both the *ab*-plane, $\xi_{ab}(T)$, and the *c*-axis, $\xi_c(T)$, exhibit a monotonic increase with rising temperature. At absolute zero (0 K), the in-plane coherence length $\xi_{ab}(0)$ is marginally greater than its out-of-plane counterpart. However, as the temperature increases and approaches the critical temperature $T_c$, the values of $\xi_{ab}(T)$ and $\xi_c(T)$ converge and become nearly identical.

To comprehensively assess orbital and spin-paramagnetic effects, we fitted the temperature-dependent $\mu_0 H_{c2}(T)$ data using the one-band, dirty-limit Werthamer-Helfand-Hohenberg (WHH) formula,[26] $\ln\frac{1}{t} = \sum_{v=-\infty}^{\infty}\left\{\frac{1}{|2v+1|} - \left[|2v+1| + \frac{\hbar}{t} + \frac{(\alpha\hbar/t)^2}{|2v+1|+(\hbar+\lambda_{so})/t}\right]^{-1}\right\}$, where $t = T/T_c$, $\hbar = (4/\pi^2)[H_{c2}(T)/|dH_{c2}/dT|_{T_c}]$, $\alpha$ is the Maki parameter, and $\lambda_{so}$ is the spin-orbit scattering constant. As shown in Figure S9, the experimental $\mu_0 H_{c2}(T)$ curves in both H$\perp$*ab* and H//*ab* orientations fit well with the WHH model in the high-temperature regime near $T_c$. In the case, it neglects the spin paramagnetic effect ($\alpha$=0) and spin-orbital interaction ($\lambda_{so}$=0), and yields out-of-plane and in-plane upper critical fields of 44.8 T and 68.7 T, respectively. The experimental $\mu_0H_{c2}(T)$ results can be well fitted by the G-L model in the current temperature range as well.

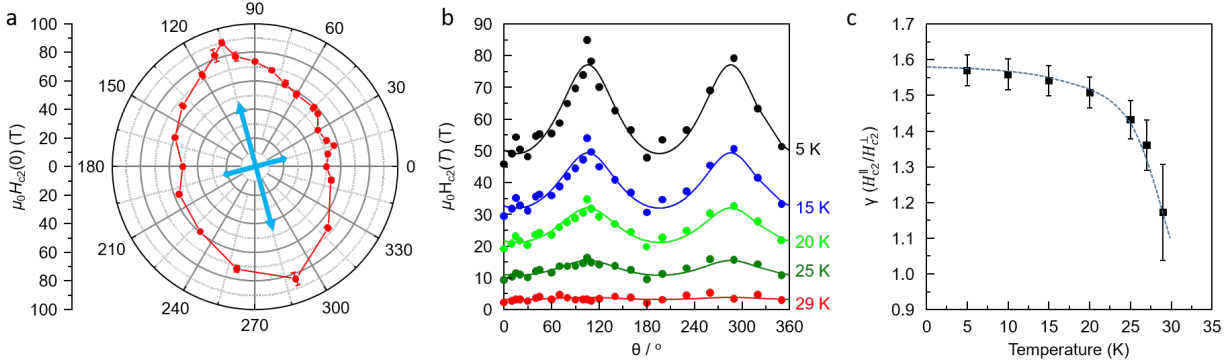

**Figure 3.** Temperature-dependent anisotropy of the upper critical field in a $Pr_4Ni_3O_{10}$ single crystal. (a) Polar plots of the angular-dependent upper critical field, $\mu_0H_{c2}(0)$ as extracted from the G-L formula. (b) Angular dependence of $\mu_0H_{c2}(T)$ at temperatures of 29 K (red), 25 K (olive), 20 K (green), 15 K (blue), and 5 K (black). Solid lines are fits based on the three-dimensional (3D) G-L model. (c) Temperature evolution of the anisotropy parameter $\gamma$.

A summary of the angular dependence of the $\mu_0H_{c2}(0)$ is presented in Figure 3a. The data exhibit clear C2 rotational symmetry, with maximum values observed at 105° and 285°, and minimum values at 15° and 195°. Angle-resolved $\mu_0H_{c2}(0)$ fitted by the WHH model further confirm the distinct twofold symmetry as demonstrated in Figure S10. To further investigate the superconducting anisotropy, the angular dependence of $\mu_0H_{c2}(T)$ was characterized at various temperatures (Figure 3b). A systematic suppression of the angular modulation amplitude in $\mu_0H_{c2}(T)$ was observed with increasing temperature. For 3D bulk superconductors, the angular dependence of $\mu_0H_{c2}$ is well-described by the G-L model: $\mu_0H_{c2}(\theta) = \mu_0H_{c2}^{\parallel}/\sqrt{cos^2\theta + \gamma^2 sin^2\theta}$, where $\mu_0H_{c2}^{\parallel}$ represents the upper critical field when the magnetic field is applied parallel to the ab-plane. Fitting the angular-dependent upper critical field data to this G-L formula yields an anisotropy parameter $\gamma$ = 1.6 at 0 K (Figure 3b). This value is consistent with the ratio $\mu_0 H_{c2}^{\parallel}(0)/\mu_0 H_{c2}^{\perp}(0)$ derived from direct experimental measurements. As shown in Figure 3c, the anisotropy parameters were determined to be 1.17±0.13 and 1.36±0.07 at 29 K and 27 K, respectively. Notably, the anisotropy decreases with increasing temperature. In contrast to two-dimensional (2D) superconductors, which typically exhibit a pronounced enhancement in anisotropy upon cooling [27-28], $Pr_4Ni_3O_{10}$ demonstrates an increase but reaching a plateau soon below the $T_c$. The weak anisotropy across different temperatures observed in $Pr_4Ni_3O_{10}$ is similar to that reported for $MgB_2$ [29], FeAs-112 [30] and FeAs-122 [31-32] system. Combined with the observed temperature dependence of the anisotropy, our results suggest that $Pr_4Ni_3O_{10}$ possesses a 3D Fermi surface topology.

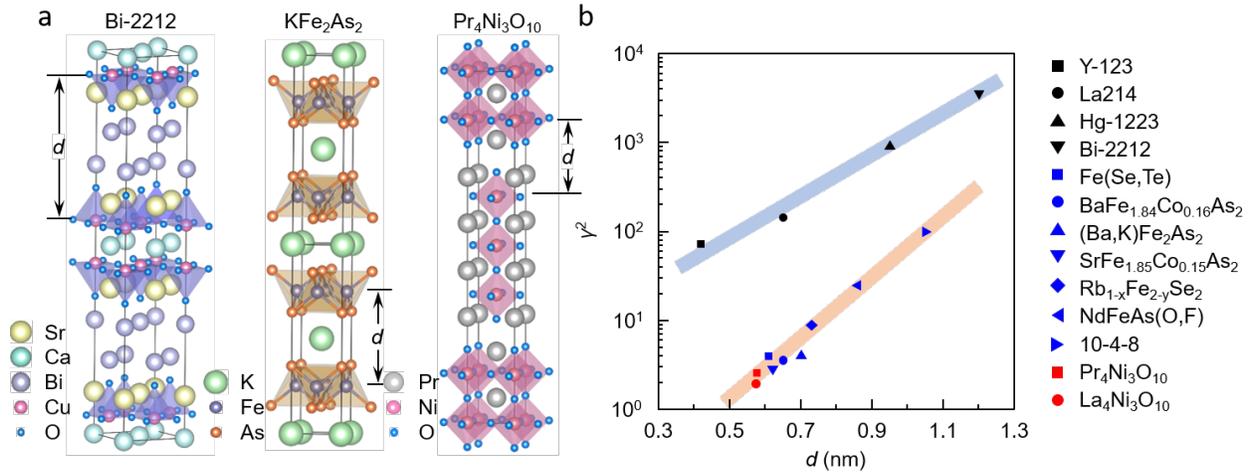

**Figure 4.** Relationship between superconducting anisotropy and interlayer coupling. (a) Schematic illustrations of the crystal structures, highlighting the interlayer distances (*d*) between the superconducting planes: $CuO_2$ layers in cuprates, FeAs/FeSe layers in FeSCs, and $NiO_2$ layers in RP-type nickelates. (b) Anisotropy parameter $\gamma^2$ plotted as a function of the adjacent interlayer distance *d* for the three families of superconductors. Data for cuprates and FeSCs are from Ref. 28, included for comparison. The red and blue shaded bands serve as visual guides to emphasize the distinct trends.

In both cuprates and FeSCs, the anisotropy parameter exhibits a strong correlation with the interlayer distance of the superconducting $CuO_2$ layers and FeAs/FeSe layers, respectively. The parameter $\gamma^2$ serves as an indicator of the coupling strength between the charge reservoir layers and the conducting superconducting layers. Additionally, it is also affected by the conductivity of the spacer layers intercalated between adjacent superconducting layers [30]. Hence, we summarized $\gamma^2$ values for cuprates, FeSCs and bulk RP-type nickelate superconductors as a function of the interlayer spacing $d$ between adjacent superconducting layers in Figure 4. Consistent with previous reports, $\gamma^2$ increases approximately linearly with $d$ in the two major high-$T_c$ superconductor families: cuprates and FeSCs. Notably, however, the two systems exhibit distinct slopes in this linear relationship, with cuprates displaying significantly higher anisotropy than FeSCs at a given interlayer distance. These differences originate from the fact that superconductivity in cuprates is predominantly confined within the $CuO_2$ layers. In contrast, FeSCs exhibit multi-orbital and 3D electronic characteristics, which lead to stronger interlayer coupling and reduced anisotropy. Surprisingly, despite sharing similar perovskite building blocks with multilayer high-$T_c$ cuprates, trilayer $Pr_4Ni_3O_{10}$ superconductors exhibit weak anisotropy comparable to that of FeSCs.

An observation was further validated by comparative studies with $La_4Ni_3O_{10}$, another bulk lanthanide RP-type nickelate with the same trilayer structure. $Pr_4Ni_3O_{10}$ incorporates $Pr^{3+}$ ions with a smaller ionic radius, [7, 16] leading to smaller lattice parameters, a higher $T_c$ (~40 K) and a higher phase transition pressure (~30 GPa) than $La_4Ni_3O_{10}$[4-5]. In contrast, $La_4Ni_3O_{10}$ undergoes a structural transition at ~15 GPa and has a $T_c$ of ~30 K. This trend is similar to that in iron-based superconductors, where the $T_c$ of $(Pr/Nd)[O_{1-x}F_x]FeAs$ (52.0-53.5 K) is twice that of $La[O_{1-x}F_x]FeAs$ (~26 K).[33-34] Electronically, $Pr_4Ni_3O_{10}$ features a two-orbital Fermi topology, while $La_4Ni_3O_{10}$ possesses a three-orbital configuration around the Fermi level. In the study of superconducting anisotropy, Pr$_4$Ni$_3$O$_{10}$ exhibits weak superconducting anisotropy with a γ of ~1.6 at 0 K that decreases with increasing temperature, while La$_4$Ni$_3$O$_{10}$ shows nearly isotropic superconductivity with γ ranging from 1.4 near $T_c$ to 1 at low temperatures. The in-plane and out-of-plane superconducting coherence lengths are both ~33 Å at 1.8 K for La$_4$Ni$_3$O$_{10}$[35]. Based on a NiO$_2$ layer thickness of 3.7 Å at room temperature (r.t.) and 49 GPa[5], we can estimate a trilayer NiO thickness of ~ 7.33 Å and trilayers distance of 5.83 Å. In contrast, Pr$_4$Ni$_3$O$_{10}$, with smaller lattice parameters, yields $\xi_{ab}(0)$ = 24 Å and $\xi_c(0)$ = 15 Å at 0 K. The thickness of superconducting NiO$_2$ trilayers is 7.27 Å and the inter-trilayer spacing is 5.77 Å at 49 GPa and 16 K (Figure S11). Thus, for both materials, the superconducting coherence lengths exceed the layer thicknesses and interlayer spacings, indicating their 3D superconducting nature and extremely low superconducting anisotropy.

Corresponding to high-pressure nickelate superconductors, there are ambient-pressure counterparts. Ambient-pressure nickelate are typically thin films, and mainly encompassed infinite-layer nickelates[36-39] and RP-type nickelates (including bilayer and hybrid-stacked structures)[40-45]. As summarized in Table S1, infinite-layer nickelate La$_{0.8}$Sr$_{0.2}$NiO$_2$ thin films exhibit substantial anisotropy near $T_c$, and gradually decreases with decreasing temperature to isotropic with anisotropy parameter of 1.5 at 2 K.[38-39] In contrast, the Nd$_{0.775}$Sr$_{0.225}$NiO$_2$ films show isotropic behavior which likely originates from orbital-selective pairing.[36, 46] Recent studies show that rare-earth ions affect the superconducting anisotropy of infinite-layer nickelate thin films *via* their 4*f* electrons, with La$^{3+}$/Pr$^{3+}$ inducing conventional layered anisotropy and Nd$^{3+}$ leading to distinct anisotropic behavior.[47] For RP-type nickelate thin films, the anisotropy parameters of La$_3$Ni$_2$O$_7$, (La,Pr)$_3$Ni$_2$O$_7$, 1212-stacked (La,Pr)$_5$Ni$_3$O$_{11}$ films and (La,Pr,Sm)$_3$Ni$_2$O$_7$ films, derived from the upper critical field at 0 K are as small as 1.4, 1.7-2.5, 1.6-2.4 and 2.1, respectively.[40-41,

[43, 48] When it comes to the trilayer nickelate bulk superconductors ($La_4Ni_3O_{10}$ and $Pr_4Ni_3O_{10}$ single crystals), the anisotropy parameter is inherently small. Consequently, it is reasonable to state that RP-type nickelates, both as thin films and bulks, show weak anisotropy. Pressure in a DAC compresses the lattice parameters of both in-plane and out-of-plane orientation, whereas thin films under biaxial compressive epitaxial strain leads to in-plane lattice compression and slight out-of-plane lattice expansion[43]. Consequently, interlayer interactions are weaker in thin films than in bulks, resulting in slightly larger anisotropy in thin films. Additionally, the distinct electronic structures between infinite-layer and RP-type nickelate superconductors may contribute to their different superconducting anisotropies.

Given that RP-type nickelates, cuprates and bismuthates all share a perovskite-related structure with $MO_6$ octahedra as building blocks, we further discuss the anisotropy of perovskite superconductors. Similar to RP-type nickelate superconductors, superconductivity in Pb/K-doped $BaBiO_3$ is exclusive to their tetragonal phase, with $T_c$ as high as 34 K. However, $BiO_6$ octahedra undergo slight tilting rather than forming perfect linear Ni-O-Ni linkages in nickelates, and it does not disrupt the 3D electronic transport network required for superconductivity. RP-type nickelate and cuprate superconductors have involved nonconventional mechanisms, while the disproportionation of Bi ion in bismuthates results possible negative U mechanism, leading to what is often called 'bipolaronic' superconductivity. Angle-resolved photoemission spectroscopy studies identified $Ba_{0.51}K_{0.49}BiO_3$ as an extraordinary BCS superconductor with an isotropic superconducting gap.[49] Cuprate superconductors exhibit considerable superconducting anisotropy due to their quasi-2D electronic structures dominated by $CuO_2$ layers while RP-type nickelate superconductors show weak anisotropy. The comparative analysis clarifies the unique position of n=3 RP nickelates in the evolution from low-dimensional cuprates to high-dimensional bismuthates superconductors systems, providing new experimental evidence for understanding the correlation between structural dimensionality and superconducting anisotropy.

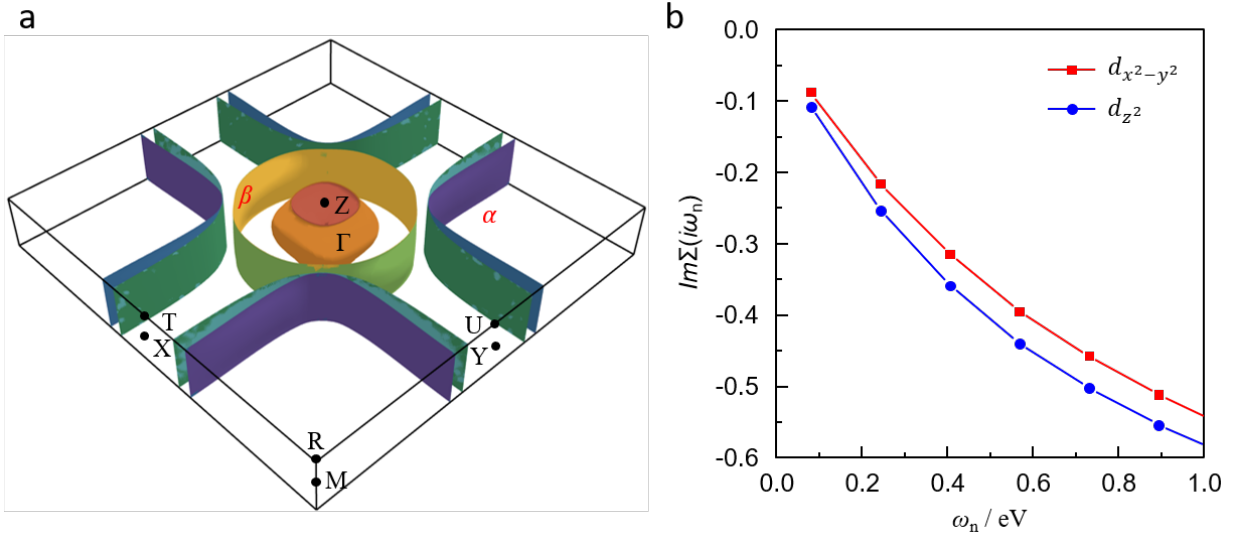

**Figure 5.** Electronic structure of the high-pressure phase of $Pr_4Ni_3O_{10}$. (a) Fermi surface calculated within the GGA+U approximation. A 3D electron pocket, originating from the bonding $d_{z^2}$ band, is observed at the Brillouin zone center (Γ point). Additional features include a hole-like $\alpha$ sheet with predominant $d_{x^2-y^2}$ orbital character near the zone corner (M point), and an electron-like $\beta$ sheet with mixed $e_g$ orbital character around the Γ point. (b) Orbital-resolved imaginary part of the self-energy, $Im\Sigma(i\omega_n)$, on the Matsubara axis at 300 K.

The anisotropy in $\mu_0 H_{c2}$ can be understood within the framework of the G-L theory. The anisotropy parameter $\gamma$ is approximately equal to the square root of the effective mass anisotropy, and can be expressed in the following equivalent forms: $\gamma = \sqrt{m_c/m_{ab}} = \xi_{ab}/\xi_c = \mu_0 H_{c2}^{\parallel}/\mu_0 H_{c2}^{\perp}$. Here, $m_c$ ($m_{ab}$), $\xi_c$ ($\xi_{ab}$) and $\mu_0 H_{c2}^{\perp}$ ($\mu_0 H_{c2}^{\parallel}$) denote the effective masses, coherence lengths and upper critical fields along $c$-axis ($ab$-plane), respectively. Given that the low-energy physics of $Pr_4Ni_3O_{10}$ can be effectively captured by two Ni $e_g$ orbitals, we propose that interactions between correlated carriers in the $3d_{z^2}$ and $3d_{x^2-y^2}$ orbitals are likely to play a significant role in superconductivity. These interactions are expected to give rise to strongly orbital-dependent anisotropic superconducting behaviors, thereby supporting an effective two-band description of system. In contrast to a 2D sheet-like topology, the GGA+$U$- calculated Fermi surface of Pr$_4$Ni$_3$O$_{10}$ exhibits significant warping effects in the $3d_{z^2}$ dominant electron pocket near the center of Brillouin zone. This warping suggests the possibility of out-of-plane circulating currents and indicates 3D electronic character (Figure 5a). To investigate the influence of electronic correlations on the anisotropic effective masses, we performed charge self-consistent DFT+eDMFT calculations. For computational simplicity, the inner- and outer-layer Ni sites were treated as identical correlated impurities in the DFT+eDMFT framework. Figure 5b displays the imaginary parts of Matsubara-frequency self-energy $Im\Sigma(i\omega_n)$ for the $3d_{z^2}$ and $3d_{x^2-y^2}$ orbitals. The $Im\Sigma(i\omega_n)$ magnitudes are larger for the $3d_{z^2}$ orbitals than for the $3d_{x^2-y^2}$ orbitals. In addition, the slope of $Im\Sigma(i\omega_n)$ at low frequencies is steeper for the $3d_{z^2}$ orbitals, indicating stronger correlation effects in the $3d_{z^2}$ orbital compared to the $3d_{x^2-y^2}$ orbital. The calculated effective mass enhancement due to electronic correlation, given by $\frac{m^*}{m} = \frac{1}{Z} = 1 - (\partial Re\Sigma(\omega)/\partial\omega)|_{\omega=0}$, is 2.46 and 2.19 for $3d_{z^2}$ and $3d_{x^2-y^2}$ carriers, respectively. The corresponding band mass anisotropy, $m_c/m_{ab}$, was found to be 1.06. Besides, it may be interesting to discuss the underlying superconducting pairing symmetry in Pr$_4$Ni$_3$O$_{10}$ based on its 3D two-orbital electronic features. Theoretical studies on its bilayer counterparts La$_3$Ni$_2$O$_7$[50] suggest that pairing originating from the $d_{z^2}$-derived γ band can favor an $s_{\pm}$-wave state with sign-reversal gaps through interband coupling between the γ band and other bands, while the pairing interaction within the $d_{x^2-y^2}$-derived bands tend to support a $d$-wave character of pairing symmetry. Since for Pr$_4$Ni$_3$O$_{10}$ both the $d_{z^2}$ band and the $d_{x^2-y^2}$ band contribute substantially to the Fermi surface with close correlation strengths, a mixed or competing $s_{\pm-}$ and $d$-wave pairing character could emerge, which deserves further experimental studies.

**Conclusion**

In summary, we have performed *in situ* high-pressure, angle-dependent electrical transport measurements on $Pr_4Ni_3O_{10}$ single crystals under applied magnetic fields. Despite sharing structural similarities with cuprates in terms of perovskite-based building blocks, $Pr_4Ni_3O_{10}$ exhibits remarkably low electronic anisotropy, comparable to that of FeSCs. To investigate the inter-unit-cell coupling and anisotropy, we employed an effective two-band model incorporating both the $3d_{z^2}$ and $3d_{x^2-y^2}$ orbitals. Our analysis underscores the essential role of interlayer coupling, mediated predominantly by the $3d_{z^2}$ orbital, in the superconductivity of RP-type nickelates. Cuprates exhibit quasi-2D behavior, whereas FeSCs display show 3D characteristics. RP-type nickelates, represented by $Pr_4Ni_3O_{10}$, bridge these two families and ultimately align more closely with the 3D behavior of FeSCs. This finding significantly refines our understanding of how structural and electronic dimensionality governs high-temperature superconductivity.

**ASSOCIATED CONTENT**

**Supporting Information**. Experimental section, temperature-dependent resistance curves, extracted $H_{c2}(\varphi)$ using the WHH relation, Arrhenius plots of the resistivities for H$\perp ab$ (a) and H//$ab$ (b), field dependence of $U_0$ for $Pr_4Ni_3O_{10}$ single crystal, lattice parameters and superconductivity anisotropy of high-temperature superconductor are supplied as Supporting Information. This material is available free of charge *via* the Internet at http://pubs.acs.org.

**ACKNOWLEDGMENT**

This work was supported by the Science and Technology Commission of Shanghai Municipality (Grant No. 25DZ3008200), National Natural Science Foundation of China (Grant Nos. 52272265, 12474018) and the National Key R&D Program of China (Grant No. 2023YFA1607400). Guang-Ming Zhang acknowledges the support from the National Key Research and Development Program of MOST of China (Grant No. 2023YFA1406400). Qiaoshi Zeng, Di Peng and Wenge Yang acknowledge the support from the National Natural Science Foundation of China (Grant Nos. 51871054, 52101187) and Shanghai Key Laboratory of Material Frontiers Research in Extreme Environments, China (Grant No. 22dz2260800), the Shanghai Science and Technology Committee, China (Grant No. 22JC1410300). Hanjie Guo acknowledges the support from the National Natural Science Foundation of China (Grant No. 12004270) and Guangdong Basic and Applied Basic Research Foundation (Grant No. 2022B1515120020). Xiangzhuo Xing acknowledges the support from the Young Scientists of Taishan Scholarship (Grant No. tsqn202408168）Zhixiang Shi acknowledges the support from the National Natural Science Foundation of China (Grant No. 12374135). The authors thank the support from Analytical Instrumentation Center (# SPSTAIC10112914), SPST, ShanghaiTech University.